# Magnetic Resonance Spectroscopy Deep Learning Denoising Using Few In Vivo Data

Dicheng Chen#, Wanqi Hu#, Huiting Liu, Yirong Zhou, Tianyu Qiu, Yihui Huang, Zi Wang, Liangjie Lin, Zhigang Wu, Jiazheng Wang, Hao Chen, Xi Chen, Gen Yan, Di Guo, Jianzhong Lin, and Xiaobo Qu*

*Abstract*— **Magnetic Resonance Spectroscopy (MRS) is a noninvasive tool to reveal metabolic information. One challenge of $^1$H-MRS is the low Signal-Noise Ratio (SNR). To improve the SNR, a typical approach is to perform Signal Averaging (SA) with *M* repeated samples. The data acquisition time, however, is increased by *M* times accordingly, and a complete clinical MRS scan takes approximately 10 minutes at a common setting *M*=128. Recently, deep learning has been introduced to improve the SNR but most of them use the simulated data as the training set. This may hinder the MRS applications since some potential differences, such as acquisition system imperfections, and physiological and psychologic conditions may exist between the simulated and *in vivo* data. Here, we proposed a new scheme that purely used repeated samples of realistic data. A deep learning model, Refusion Long Short-Term Memory (ReLSTM), was designed to learn the mapping from the low SNR time domain data (24 SA) to the high SNR one (128 SA). Experiments on the *in vivo* brain spectra of 7 healthy, 2 tumor and 1 cerebral infarction region shows that only using 20% repeated samples, the denoised spectra by ReLSTM could provide comparable estimated concentrations of metabolites to 128 SA. Compared with the state-of-the-art low-rank denoising method, the ReLSTM achieved a lower relative error and the Cramér-Rao lower bounds in quantifying some important biomarkers. In summary, ReLSTM can perform high-fidelity denoising of the spectra under fast acquisition (24 SA), which would be valuable to MRS clinical studies.**

*Index Terms*— **Magnetic resonance spectroscopy, deep learning, denoising, *in vivo* data, metabolites.**

## I. Introduction

MAGNETIC Resonance Spectroscopy (MRS) is a non-invasive imaging technique [1-4] to measure the metabolites and provide the quantitative analysis [5-8] of metabolites in brains. For example, $^1$H-MRS can detect the important metabolic biomarkers in human brain diseases [9]:

- N-acetylaspartate (NAA) indicates the abundance and viability of neurons. Damage [10] or tumor [11] of the brain will decrease the NAA;
- Choline (Cho) is one of the components of membrane phospholipid metabolism and is elevated in brain tumor [11];
- Creatine (Cr) reflects cerebral energy metabolism and its ratios with other metabolites will significantly change in some diseases. For example, Cho/Cr increases in tumor [12], and Lactate/Cr increases in cerebral infarction [13];
- Glutamate (Glu) is an intracellular neurotransmitter marker and increases in cerebral ischemia and hypoxia [14];
- Myo-Inositol (mI/Ins) is thought to be the metabolite of hormone-sensitive nerve receptors and its levels correlate with cerebral astrocytoma grade [15].

However, the low SNR and narrow spectral dispersion for the *in vivo* brain $^1$H spectra from the limited magnetic field of the clinical MR scanner, lead to difficulties in further metabolic quantification and analysis [5][16]. We aimed to improve the low SNR of the spectra in this paper. Although the SNR may be affected by many factors, such as magnetic field inhomogeneity [17], subject physiological motion [18] and hardware defects [19], the main cause is that the signal intensity is very low due to the low concentration of metabolites [20], excited at the limited magnetic field of the clinical MR scanner.

To gain a relatively high SNR with the limited magnetic field of the clinical MR scanner, MRS scanning usually requires repeated sampling, and then averaging these repeated samples. This process is called Signal Averaging (SA) [18][21] which is named differently in scanner vendors, e.g. NSA in Philips, NEX in GE and Average in Siemens. Generally, within 4 to 5 minutes, one can acquire single-voxel MRS under the default setting (128 SA, TR=2000 ms). But in clinical practice, spectrum acquisition is an extra process of anatomy imagining and MRS only needs when it is necessary, e.g. brain tumor diagnosis. Thus, about 10 minutes may be required for the complete imaging process (anatomy imaging, shimming, correction and two signals with and without water suppression for MRS calibration and quantification). If the SA (the number of averages) is increased to get higher SNR, the scanning time will be longer than 10 minutes, which may make subjects feel uncomfortable and introduce more motions. Thus, how to improve the SNR as much as possible while reducing the *M* is important for fast MRS.

Denoising the spectra from fewer repeated samples is a

This work was supported in part by National Natural Science Foundation of China (62122064, 61971361, 61871341, 61811530021), Natural Science Foundation of Fujian Province of China (2021J011184), Health-Education Joint Research Project of Fujian Province (2019-WJ-31), Xiamen University Nanqiang Outstanding Talents Program.
D. Chen, W. Hu, H. Liu, Y. Zhou, T. Qiu Y. Huang, Z. Wang and X. Qu are with Department of Electronic Science, Fujian Provincial Key Laboratory of Plasma and Magnetic Resonance, Xiamen University, Xiamen 361005, China (*corresponding author with Email: quxiaobo@xmu.edu.cn).
L. Lin, J. Wang and Z. Wu are with Philips, Beijing 100016, China.
H. Chen is with School of Biomedical Engineering, Shanghai Jiao Tong University, Shanghai 200240, China.
X. Chen is with the McLean Hospital, Harvard Medical School, Belmont, MA 02478, USA.
G. Yan is with Department of Radiology, The Second Affiliated Hospital of Xiamen Medical College, Xiamen 361021, China.
D. Guo is with School of Computer and Information Engineering, Xiamen University of Technology, Xiamen 361024, China.
J. Lin is with Department of Radiology, The Zhongshan Hospital affiliated to Xiamen University, Xiamen 361004, China.

potential strategy to improve the SNR. So far, many optimization methods have been proposed for denoising [22-28]. Among them, the low-rank method is a powerful approach [22-23][27-28] as its hypothesis, that the time domain signal of MRS is exponential functions, fits the theory of magnetic resonance physics very well [22][29]. The exponential assumption, however, may deviate from real data due to field inhomogeneity or other imaging system imperfections [30-31].

Recently, Deep Learning (DL), a representative artificial intelligence technique, has been introduced to the magnetic resonance imaging, e.g. image reconstruction [32-37] and magnetic resonance fingerprinting [38-41]. DL has also shown great potential by its end-to-end nonlinear fitting in MRS [42-43], e.g. fittings [44-46], reconstruction [47-49], denoising [50] and artifact removal [44][51].

The training of a DL network commonly requires a large amount of data. For MRS, acquiring many human spectra, particularly high SNR spectra as training labels, is highly time-consuming and difficult. Most MRS DL methods generated synthetic training data through physical evolution or analytical models [44-45][47][52-55], which have been recently reviewed in [52]. Spectral parameters, under a reasonable biomedical range, have been used to generate training data and have achieved good performance [44-45]. However, if the spectral parameters are out of the pre-defined range or some other factors, e.g. residual water peaks and lesion effects, are not modelled accurately, the DL model may need to be retrained. To enhance the adaptivity to real data under synthetic data training, traditional signal priors, e.g. spatial smoothness, have been successfully applied to improve the DL MRS denoising but at the cost of extra regularization and more hyperparameters [47].

Furthermore, some potential differences, such as acquisition system imperfections, physiological and psychological conditions, may exist between the simulated training set and realistic data, which may hinder the applications of DL in MRS. Thus, few studies [46][50] tried to use realistic data to train DL models. In the application of spectrum fitting [46], partial multi-voxel brain data from 10 subjects have been selected (spectral width of peaks that are greater than 18Hz) to learn the spectral parameters in an unsupervised manner. During the review process of our work, another research [50] tried to learn the network mapping from low SNR (SA=8) to high SNR (SA=192) on phantom data and then apply the network to real brain MRS, improving the SNR by 47%. However, the mean square error of denoising on real brain MRS is 2 times larger than that of phantom data, indicating the reduced performance due to the transferring from phantom to real data.

In this work, a supervised deep learning denoising approach was proposed to combine different repetitions of *in vivo* human brain single-voxel MRS data to make a huge realistic training set. A large number of training data can be obtained by randomly selecting $m$ repeated samples from the $M$ repetitions. This number is a huge combination number. For example, if we randomly select 24 from $M$=128 repetitions, the combination number is in the order of $10^{25}$.

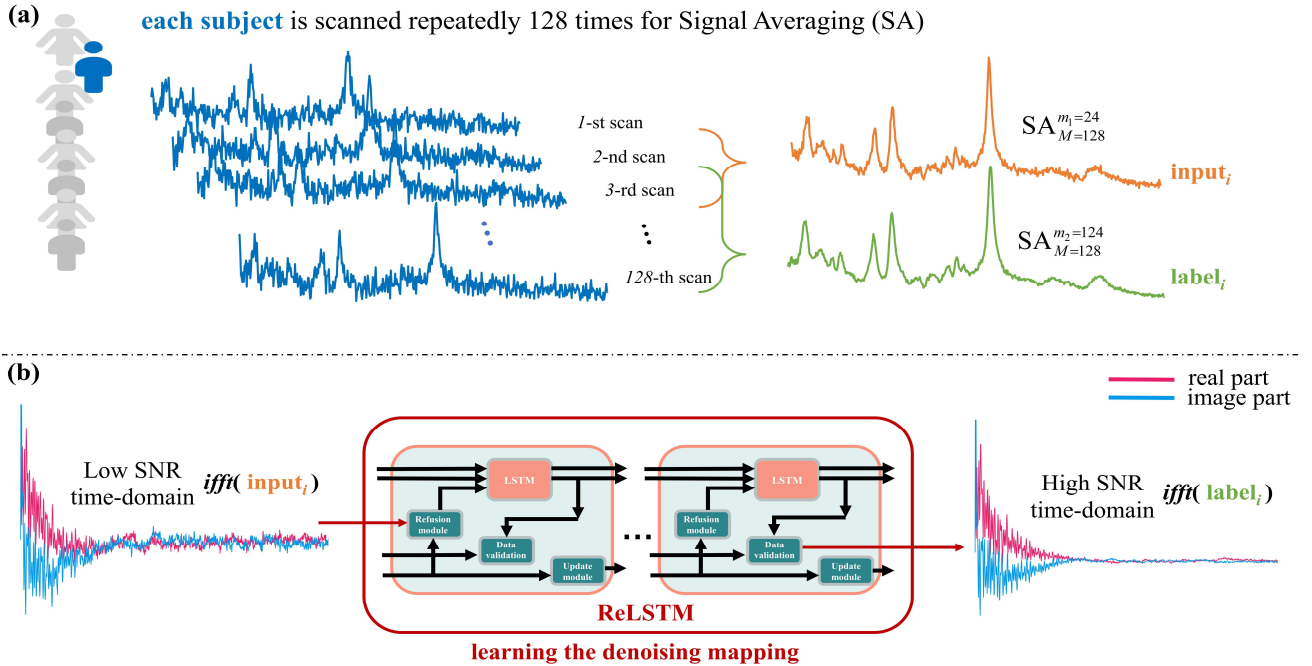

**Fig. 1.** The entire training phase of the proposed ReLSTM. (a) one training input was the average of $m_1$=24 repeated samples which were randomly selected out of total repeated scans $M$=128. For convenience, $SA_M^m$ was denoted as this averaged spectrum, which was called $m$ SA. Then, one corresponding training high SNR label was $SA_{M=128}^{m_2=124}$ (124 SA); (b) the proposed denoising ReLSTM was trained for fitting the mapping between the low SNR time domain of MRS *ifft*(input$_i$) and the corresponding high SNR time domain of MRS *ifft*(label$_i$) where *ifft* represents inverse Fourier transform.

Besides, different from most DL MRS denoising methods that are performed in the frequency domain [44-45][50], the proposed method provides an alternative way to remove noise in the time domain. Specifically, a Long Short-Term Memory (LSTM) network [56] is modified to mine the correlation among time domain data points. Recently, the LSTM has been successfully applied to nuclear magnetic resonance spectra for fast sampling reconstruction [57]. Further, we proposed a new framework that embedded the refusion mechanism and data validation into the LSTM to improve the MRS denoising (Section III-B). The network learns the denoising mapping from 24 repetitions (low SNR) to 124 repetitions (high SNR) (Fig. 1(b)), which is similar to the magnetic resonance fingerprinting [38-41] that learns the T1 or T2 mapping from signal repetition.

The main contributions of this work include:
1) Build *in vivo* training data by randomly recombining its repeated samples.
2) Propose a new improved LSTM neural network to suit MRS time domain denoising trained by a few measured data.
3) Save about 80% ( down from about 5 minutes to 1 minute) single-voxel spectra acquisition time with comparable metabolic concentrations Relative Error (RE) and Standard Deviations Percentage using Cramér-Rao lower bounds percentage (SDP) to 128 SA.
4) Verify denoising performance on MRS acquired from multiple brain regions of both healthy and patient subjects.

The rest of the paper was arranged as follows. Section II introduced the acquisition of *in vivo* data and the generation of the training set. Section III proposed the network structures and provided implementation details. Section IV quantitatively compared the proposed method with 128 SA and the state-of-the-art low-rank method. Finally, the selection of hyperparameters and limitations were discussed in Section V.

## II. DATA PREPARATION

### A. In Vivo MRS Acquisition

The experimental single-voxel $^1$H-MRS scans were performed on 21 volunteers, including 18 healthy subjects and 3 patients. Data acquisition was approved by the institutional review board of the Zhongshan Hospital affiliated to Xiamen University. All measurements were conducted on the Philips 3T DNA scanner with a 32-channel head coil, using a point resolved spectroscopy (PRESS) sequence [58] with parameters: TR/TE=2000ms/35ms, bandwidth=2kHz, dwell time=0.0005s, NPTS (number of points) = 2048, number of averaging =128, voxel size = 20mm×20mm×20mm. Each complex raw time domain data was in the dimension of 2048×128.

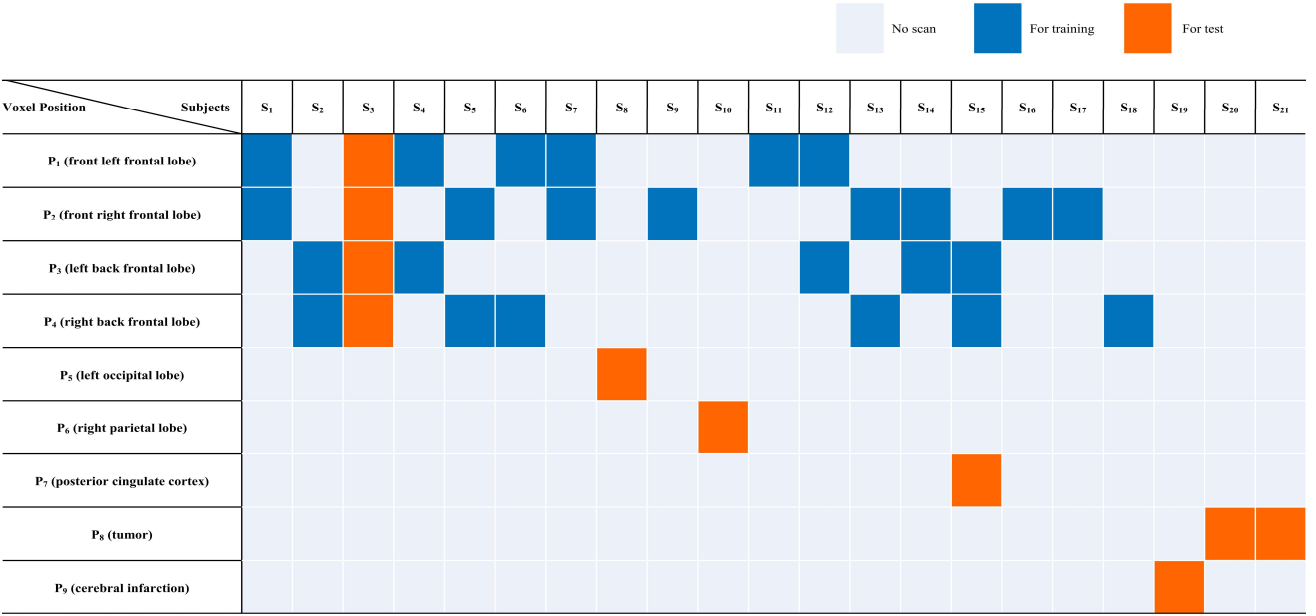

**Fig. 2.** The arrangements of *in vivo* single-voxel $^1$H-MRS. $S_i$ represents the *i-th* subjects and $P_j$ is for the *j-th* voxel position. The following descriptions uses $S_i$ $P_j$ to represent the spectra of the *i-th* subjects at the *j-th* voxel position.

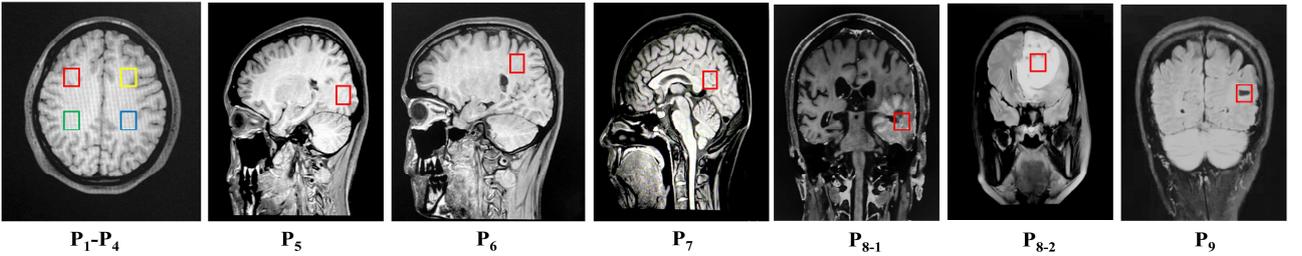

**Fig. 3.** Voxel positions of subjects. Among them, $P_1$-$P_4$ are the frontal lobes in the front-right (red), front-left (yellow), back-right (green) and back-left (blue), respectively; $P_5$ indicates the left occipital lobe; $P_6$ is the right parietal lobe; $P_7$ is the posterior cingulate cortex; $P_{8-1}$ and $P_{8-2}$ are tumor regions in two different patients; $P_9$ is the cerebral infarction region.

A total of 35 *in vivo* measured MRS data were collected from 18 healthy subjects' brains ($S_1$-$S_{18}$), 1 cerebral infarction ($S_{19}$) and 2 patients' tumor regions ($S_{20}$-$S_{21}$). Fig. 2 shows the detailed arrangement of the MRS collection of each subject and Fig. 3 shows the corresponding voxel positions. Specifically, we only chose the relatively stable spectra of the frontal lobe ($P_1$-$P_4$) for training. In the feasibility testing, we not only chose the spectra of the frontal lobe ($S_3P_{1-4}$), but also of other positions such as the occipital lobe, parietal lobe, cingulate cortex ($P_5$-$P_7$), and even the lesion regions ($P_8$-$P_9$) from different subjects. These test spectra were processed as the 24 SA denoising inputs with SNR ranging from 7.8 to 26.0.

*B. Data Augmentation*

We generated training pairs by randomly selecting different numbers of repeated samples in the complex raw MRS data. In detail, for each raw data that has *M* repeated samples, one can randomly be selected $m_1$ spectra ($m_1$<*M*) from them and then average the $m_1$ spectra as a new spectrum with a specific SNR. The SNR increases as $m_1$ increases and the number of combinations is $C_M^{m_1}$.

In our case, we randomly selected $m_1$=24 repeated samples out of total repeated scans *M*=128 and averaged them as one low SNR input$_i$ = $\text{SA}_{M=128}^{m_1=24}$ (24 SA). A corresponding high SNR label$_i$ = $\text{SA}_{M=128}^{m_2=124}$ (124 SA) was obtained by averaging 124 repeated samples (Fig.1(a)). Theoretically, each raw data can generate $C_{128}^{24} \times C_{128}^{124}$ training pairs (input$_i$, label$_i$). Following this data augmentation, 1000 items (input$_i$, label$_i$) with *i*=1,2,3,4,…,1000, were generated from each raw MRS data. A total of 25 healthy frontal lobe spectra (blue grids in Fig. 2) were selected as the training set and we consequently generated 25000 (25×1000) items for training. In addition, the testing set (marked with orange blocks in Fig. 2) was also generated as 100 items per raw MRS. Consequently, 1000 (10×100) spectra were obtained for the test.

Finally, each training pair (input$_i$, label$_i$), *i*=1,2,3,4,…, 25000 would be transformed into the time domain of MRS (Fig. 1(b)).

III. METHODS

In this section, we first introduced the basic LSTM network into MRS time-domain denoising. Then, we proposed a new LSTM denoising scheme with the Refusion module and Data validation. Finally, implementation details were provided.

*A. MRS Denoising with Basic LSTM*

The LSTM are well-suited to classification and prediction for time series. In this work, we adopted the LSTM to denoise the time signals of MRS since these signals are also time series. To our best knowledge, it is the first time to apply the LSTM to the time domain MRS denoising.

The structure of the LSTM consists of an LSTM cell and the output layer (Fig. 4(c)). The main idea of LSTM denoising is to use the low SNR sliding fragment $\mathbf{x}'_{t-1}$ from the previous time step *t*-1 to predict the current denoised point $\hat{x}_t$:

$$\hat{x}_t = \text{LSTM}(\mathbf{x}'_{t-1}|\boldsymbol{\theta}), \qquad (1)$$

where $\boldsymbol{\theta} = \{\boldsymbol{\theta}^h, \boldsymbol{\theta}^{lr}\}$ represents trainable parameters, including the LSTM cell $\boldsymbol{\theta}^h$ and output layer $\boldsymbol{\theta}^{lr}$.

By using the input sliding fragment $\mathbf{x}'_{t-1}$, the LSTM cell updates the LSTM states from ($\mathbf{c}_{t-1}$, $\mathbf{h}_{t-1}$) to ($\mathbf{c}_t$, $\mathbf{h}_t$), in which the $\mathbf{h}_t$ will be used for denoising in the output layer. The data flow in the LSTM is divided into the following steps:

1) Concatenate the previous fragment $\mathbf{x}'_{t-1}$ of the input series and the $\mathbf{h}_{t-1}$ which passes down from the previous network hidden state. Then, the concatenation result $\left[\mathbf{x}'_{t-1}, \mathbf{h}_{t-1}\right]^T$ multiplies the trainable matrices $\mathbf{W}^d, \mathbf{W}^i, \mathbf{W}^f, \mathbf{W}^o$, respectively. Then, the four LSTM cell intermediate variables can be expressed as below:

$$\begin{cases} \mathbf{z}^d = \tanh(\mathbf{W}^d \left[\mathbf{x}'_{t-1}, \mathbf{h}_{t-1}\right]^T + \mathbf{b}^d) \\ \mathbf{z}^i = \sigma(\mathbf{W}^i \left[\mathbf{x}'_{t-1}, \mathbf{h}_{t-1}\right]^T + \mathbf{b}^i) \\ \mathbf{z}^f = \sigma(\mathbf{W}^f \left[\mathbf{x}'_{t-1}, \mathbf{h}_{t-1}\right]^T + \mathbf{b}^f) \\ \mathbf{z}^o = \sigma(\mathbf{W}^o \left[\mathbf{x}'_{t-1}, \mathbf{h}_{t-1}\right]^T + \mathbf{b}^o) \end{cases}, \quad (2)$$

where $\mathbf{z}^i$, $\mathbf{z}^f$ and $\mathbf{z}^o$ are achieved the input gate, forget gate and output gate control through the sigmoid function $\sigma$ in the range of [0,1] respectively. According to the tanh function, the $\left[\mathbf{x}'_{t-1}, \mathbf{h}_{t-1}\right]^T$ is converted to the network features $\mathbf{z}^d$ in the range of [-1,1]. The $\mathbf{b}^d, \mathbf{b}^i, \mathbf{b}^f, \mathbf{b}^o$ are the bias for each gate control.

2) Update the next time step LSTM cell memory state $\mathbf{c}_t$ according to:

$$\mathbf{c}_t = \mathbf{z}^f \odot \mathbf{c}_{t-1} + \mathbf{z}^i \odot \mathbf{z}^d, \qquad (3)$$

where $\odot$ is for Hadamard product.

3) Update the next time step LSTM cell hidden state $\mathbf{h}_t$ as:

$$\mathbf{h}_t = \mathbf{z}^o \odot \tanh(\mathbf{c}_t). \qquad (4)$$

Consequently, the whole process in LSTM cell is:

$$(\mathbf{h}_t, \mathbf{c}_t) = \text{LSTMcell}(\mathbf{x}'_{t-1}, \mathbf{h}_{t-1}, \mathbf{c}_{t-1}|\boldsymbol{\theta}^h), \qquad (5)$$

where $\boldsymbol{\theta}^h$ includes $\{\mathbf{W}^d, \mathbf{W}^i, \mathbf{W}^f, \mathbf{W}^o, \mathbf{b}^d, \mathbf{b}^i, \mathbf{b}^f, \mathbf{b}^o\}$.

4) Predict the current denoised point $\hat{x}_t$ by using the updated $\mathbf{h}_t$ in the output layer with linear regression:

$$\hat{x}_t = \text{Output layer}(\mathbf{h}_t|\boldsymbol{\theta}^{lr}). \qquad (6)$$

5) Repeat to call the LSTM cell and output layer, following Eqs. (2)-(4) and (6), to predict the next denoised point $\hat{x}_{t+1}$ by

using the next time step sliding fragment $\mathbf{x}'_t$, until the sliding fragment reaches the end of the input series.

In our work, empirical settings were as follows: The $r$ ($r$=4) is the size of the fragment and the fragment is the $\mathbb{R}^{2r}$ vector (splice the imaginary part after the real part). The state $\mathbf{h}$ or $\mathbf{c}$ was with the size of $\mathbb{R}^{1024}$. Each trainable parameter used Gaussian initialization with the size of $\mathbb{R}^{1024\times(1024+2r)}$ in $\mathbf{W}^d, \mathbf{W}^i, \mathbf{W}^f, \mathbf{W}^o$ and $\mathbb{R}^{1024}$ in $\mathbf{b}^d, \mathbf{b}^i, \mathbf{b}^f, \mathbf{b}^o$. The output layer was a fully connected linear regression ($\mathbf{W}^r \in \mathbb{R}^{2\times1024}, \mathbf{b}^r \in \mathbb{R}^2$) for predicting the denoised point that contains the real and imaginary parts of $\hat{x}_t \in \mathbb{C}^r$. Finally, we fed the low SNR time domain of MRS (24 SA) into the LSTM to learn the denoised label (124 SA).

*B. MRS Denoising with Refusion LSTM*

A limitation of the basic LSTM is that it may weaken the signal intensity at the beginning of the early time decay when denoising the spectra that have multiple spectral peaks. A toy example of denoising spectrum with multiplet (Fig. 5) shows that the early decay was weakened after LSTM denoising, resulting in compromised spectral peaks.

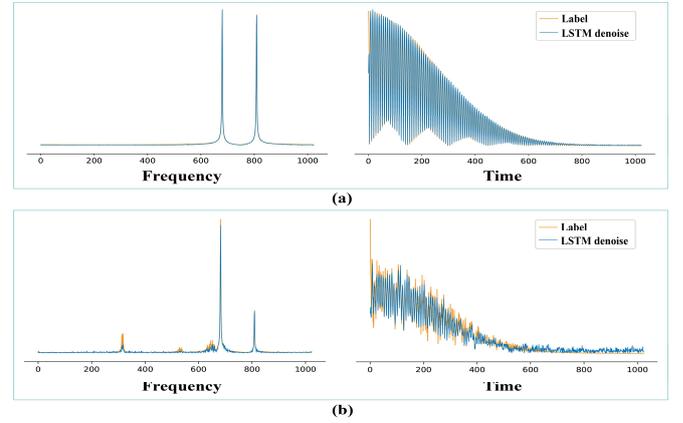

**Fig. 5.** The denoising results of simulated spectra by the LSTM and the corresponding time domain signals. (a) Spectrum with singlets; (b) Spectrum with multiplets.

To compensate for this weakening, we first set the data validation so that the denoised point not only considered the current output of the LSTM $\hat{x}_t$, but also the intensity of the current input point $x_t \in \mathbb{C}^1$. Second, the denoised outputs $\mathbf{x}_{t-1}^{out} = [x_{t-r}^{out}, x_{t-r+1}^{out}, ..., x_{t-1}^{out}] \in \mathbb{R}^{2r}$ were explored as a priori knowledge feeding into the model.

With these two new inputs $\mathbf{x}_{t-1}^{out}$ and $x_t$ (Fig. 4(a)), the ReLSTM denoised output point $x_t^{out} \in \mathbb{C}^1$ could be expressed as:

$$x_t^{out} = \text{ReLSTM}(\mathbf{x}_{t-1}, \mathbf{x}_{t-1}^{out}, x_t | \boldsymbol{\theta}), \quad (7)$$

where $\mathbf{x}_{t-1} = [x_{t-r}, x_{t-r+1}, ..., x_{t-1}] \in \mathbb{R}^{2r}$ is a size $r$ sliding fragment of the input time domain data and the $\boldsymbol{\theta}$ is a set of trainable parameters in the network. The interior of the LSTM was also upgraded by the refusion module, data validation and update module (Fig. 4(b)). The data flow conduction was much similar to the LSTM, but some new flows were added (red arrows in Fig. 4). Details are as follows:

1) The refusion module was used to mix the two sliding fragments $\mathbf{x}_{t-1}^{out}$ and $\mathbf{x}_{t-1}$ into the LSTM input fragments $\mathbf{x}'_{t-1}$ by three linear layers and one *Relu* function.

$$\mathbf{x}'_{t-1} = f_{lrl}(Relu(f_{lrl}(f_{lrl}((\mathbf{x}_{t-1}^{out}, \mathbf{x}_{t-1})|\boldsymbol{\theta}_1^{lrl})|\boldsymbol{\theta}_2^{lrl}))|\boldsymbol{\theta}_3^{lrl}), \quad (8)$$

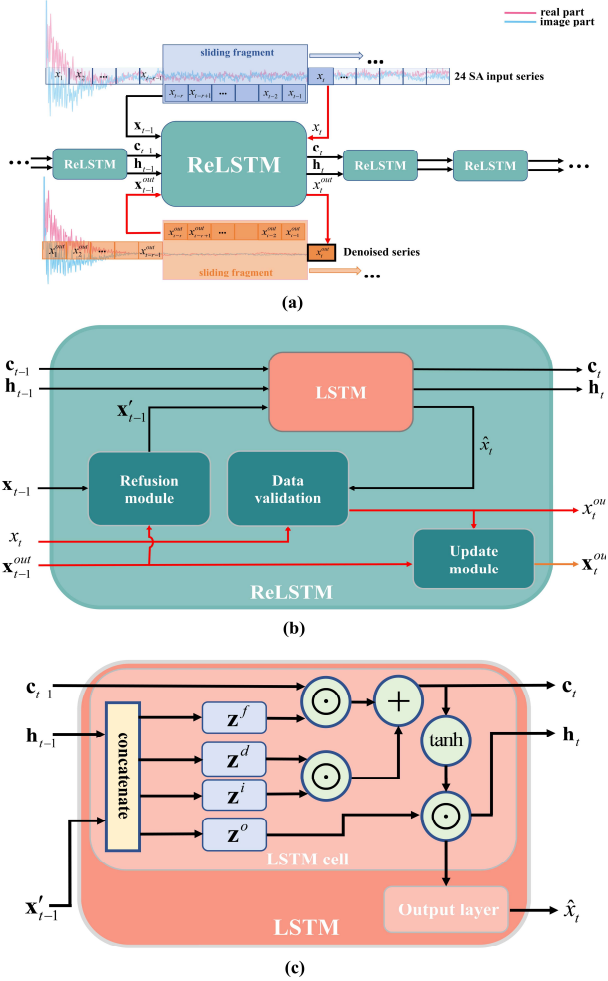

**Fig. 4.** The data flow conduction of the ReLSTM at the timestep *t*-1. (a) The 24 SA sliding fragment $\mathbf{X}_{t-1}$, corresponding sliding fragment of previous network outputs $\mathbf{X}_{t-1}^{out}$ and current 24 SA series point $x_t$ are as the ReLSTM input to predict the next time step *t* denoised point $x_t^{out}$; (b) For each time step, the data flow from input to output runs successively through the refusion module, LSTM including the LSTM cell and output layer, and data validation; (c) The previous states $\mathbf{h}_{t-1}$, $\mathbf{c}_{t-1}$ and fragment $\mathbf{x}'_{t-1}$ are fed into the LSTM cell to update its states $\mathbf{h}_t$ and $\mathbf{c}_t$. And then the updated $\mathbf{h}_t$ is used to predict the LSTM denoised point $\hat{x}_t$. Note: the red arrows represent the new data flow of the ReLSTM that are not exist in the basic LSTM.

where $\boldsymbol{\theta}_i^{lrl}, i = 1, 2, 3$ represents the training parameters of each linear layer $f_{lrl}$.

2) The $\mathbf{x}'_{t-1} \in \mathbb{R}^{2r}$ were fed into the LSTM cell for updating $\mathbf{h}_{t-1}, \mathbf{c}_{t-1}$ by using Eqs. (2)-(4).

3) The output layer was applied with the linear regression to predict an LSTM denoised point $\hat{x}_t \in \mathbb{C}^1$ by using $\mathbf{h}_t$ which is the same as the Eqs. (6).

4) The data validation was the fidelity term for balancing the current input signal point $x_t$ and the LSTM denoised output $\hat{x}_t$ which is expressed as:

$$x_t^{out} = f_{dv}\left(\hat{x}_t, x_t \middle| \boldsymbol{\theta}^{dv}\right), \quad (9)$$

where $x_t^{out} \in \mathbb{C}^1$ is the final ReLSTM denoised output of time step $t$ and $\boldsymbol{\theta}^{dv}$ represents training parameters of data validation.

5) The update module was for adding the $x_t^{out}$ into new $\mathbf{x}_{t+1}^{out} = [x_{t-r+1}^{out}, x_{t-r+2}^{out}, ..., x_t^{out}]$ to the denoising in the next time step.

6) Both sliding fragments $\mathbf{X}_{t-1}$ and $\mathbf{X}_{t-1}^{out}$ would slide next one step at the same time to capture the new series fragments $\mathbf{X}_t$ and $\mathbf{X}_t^{out}$. Repeated 1) - 6), until they slid to the last point of the input series.

Finally, the network parameters are shown in **TABLE I**.

**TABLE I**
Parameters of Network Layers

|  | Layers (input size, output size) |
|---|---|
| Refusion module | Linear ($r \times 2 \times 2, r \times 2$) + Linear ($r \times 2, r \times 32$) + Relu + Linear ($r \times 32, r \times 128$) |
| LSTM cell | LSTM cell ($r \times 128, 1024$) |
| Output layer | Linear (1024, 2) |
| Data validation | Linear (2+2, 4) + Relu + Linear (4, 2) |

Note: The layer description follows the standard in PyTorch 1.4. The size of $r$ is set by 4.

*C. Implementation*

The network was implemented by the PyTorch. Model parameters $\boldsymbol{\theta} = \{\boldsymbol{\theta}^{lrl}, \boldsymbol{\theta}^h, \boldsymbol{\theta}^{lr}, \boldsymbol{\theta}^{dv}\}$ were trained by an Adam optimizer with the learning rate $10^{-3}$ and the batch size 128 to minimize the Mean Absolute Error loss function of the time series. The training was carried out on one NVIDIA RTX2080Ti GPU for 50 epochs. Before training, all the signal was normalized with their max value in the time domain.

## IV. RESULTS

*A. Compared Method*

The low-rank Hankel denoising method (LR) assumes that the noise-free MRS time domain signal is a linear combination of a few exponential functions, resulting in the Hankel matrix converted from the time domain signal to be low rank [22][59]. Since noise will be of full rank, thus noise could be suppressed with a low-rank regularization as:

$$\min_{\mathbf{x}} \|\mathbf{R}\mathbf{x}\|_* + \frac{\lambda}{2}\|\mathbf{y} - \mathbf{x}\|_2^2, \quad (10)$$

where $\|\cdot\|_*$ denotes the nuclear norm defined as the sum of the singular values and this norm is an indicator of low-rankness [22][59]. The $\mathbf{R}$ is the operation that transforms a vector into a Hankel matrix $\mathbf{R}\mathbf{x}$. The $\lambda$ is a parameter that balances the nuclear norm and the data fidelity between a denoised $\mathbf{x}$ and the noisy time domain signal $\mathbf{y}$. An improved low-rank denoising CHORD [28] is selected as the compared LR method which is automatically set $\lambda$ by the estimated noise level.

In addition, we used 24 SA as the low SNR denoising input and 128 SA as the high SNR reference in our experiments.

*B. Assessment indicators*

- *SNR*

The denoising performance is evaluated by the SNR [60]:

$$\text{SNR} = \frac{\max(fft(\mathbf{x}) - fft(\mathbf{x})_{\text{baseline}})}{2\sqrt{\frac{1}{\text{NPTS}}\|fft(\mathbf{x}) - fft(\mathbf{x})_{\text{fitted}}\|_2^2}}, \quad (11)$$

where $\|\cdot\|_2^2$ denotes the $l_2$ norm of a vector, $fft(\mathbf{x})$ is a denoised spectrum by performing Fourier transform on the denoised time domain signal $\mathbf{x}$. $fft(\mathbf{x})_{\text{baseline}}$ and $fft(\mathbf{x})_{\text{fitted}}$ denote the fitted baseline and spectrum, respectively. Note that all the fittings in this paper were conducted with LCModel [60].

- *Concentration Accuracy*

We selected 5 metabolites that are commonly used as biomarkers of brain disease, including tNAA (NAA+NAAG), tCr (Cr+PCr), tCho (GPC+PCh), Glx (Glu+Gln) and mI/Ins in concentration evaluation. The average correlation coefficient between Glu and Gln is about -0.3 (at the boundary of the threshold -0.3 provided by references [20][60]) of our new *in vivo* data. Since the Glu and Gln cannot be well discriminated at 3T, we evaluated Glx (the summation of Glu and Gln) instead of the individual of Glu and Gln. Following the same idea that the summation is much more accurate than the individual concentrations [60], tNAA (NAA+NAAG), tCr (Cr+PCr) and tCho (GPC+PCh) were used for other concentration evaluations.

The tCho, tNAA, tCr, Glx and mI concentrations in each spectrum were estimated by LCModel using "mmol/kg" as the unit and quantified parameters details of LCModel processing were described in Supplement Fig. S1. In the relative concentration, when tCr or tNAA changes significantly in patients, tCho is used as the reference concentration, otherwise used tCr [18][60].

The concentration accuracy was evaluated by the relative error (RE) which measures the difference between the concentration $a_{128\text{SA}}$, that was estimated on a high SNR reference (128 SA), and those of the denoised spectrum, $a_{\text{denoised}}$ as:

$$\text{RE} = \frac{|a_{128\text{SA}} - a_{\text{denoised}}|}{a_{128\text{SA}}} \times 100, \quad (12)$$

where $|\cdot|$ is the absolute value operation. A failure of concentration estimation is claimed if the RE is larger than 20 [61].

- *Reliability Evaluation*

The reliability of concentrations was indicated by the Standard Deviations Percentage (SDP) which was defined as the standard deviation by Cramér-Rao lower bounds [60]. The SDP ranges from 0 to 999 and SDP ≤ 15 is a standard of acceptable reliability in LCModel [60]. Lower SDP means a more reliable concentration estimation. A failure of concentration reliability is claimed if the SPD is larger than >15 in this paper.

- *Error Bar*

For describing the error bar of the SNR, estimated concentration and SDP, the confidence interval with $\alpha = 0.99$ was applied as:

$$\bar{x} \pm z_{\frac{\alpha}{2}} \frac{S}{\sqrt{N}}, \alpha = 0.99, \tag{13}$$

where $\bar{x}$ and $S$ is the average and sample variance of $N$ measurements, $z_{\frac{\alpha}{2}}$ is the standard normal distribution statistics.

### C. Experiments

In the SNR improvement analysis, we performed the LR as well as ReLSTM denoising on the 24 SA spectra in the 1000 test spectra (SNR range from 7.8 to 26.0). The SNR improvement by denoising is shown in Fig.6. Overall, the ReLSTM achieved a higher SNR (36.3) than the 128 SA reference (30.0), which was slightly lower than the LR (37.8). Specifically, the SNR improvement of the proposed ReLSTM in the 10 *in vivo* regions was all higher than 128 SA, which indicated the effectiveness of ReLSTM denoising.

In the denoising reliability analysis, we set 128 SA spectra as the high SNR references and evaluated the quantification RE between the denoised results of the ReLSTM and LR and the references. In addition, we also set RE ≤ 20 and SDP ≤ 15 as the criteria of acceptable reliability and counted the percentage of those which are exceeding the criteria (failure rate (FR)). As can be seen in TABLE II, validated by 1000 *in vivo* data, the denoised spectra by the ReLSTM could provide comparable and reliable absolute and relative concentration estimations for 5 major brain metabolic concentrations to 128 SA.

Specifically, in all absolute concentration quantification (Total in TABLE II), the REs did not exceed 10.0% (Glx) and the FR (RE>20) was no more than 8.7% (mI). The minimum error could reach 3.2 ± 0.2 at the level of mmol/kg in the tNAA. Also, the SDP did not exceed 6.7 with the FR (SDP>15) = 6.7%. In the MRS of the healthy frontal lobe ($P_1$-$P_4$), even the most challenging case, Glx, achieved RE of 10.1 ± 1.0 and corresponding FR of 7.8%. All other metabolic quantifications achieved RE ≤ 4.6 with FR = 0%. Also, the low SDP ≤ 6.5 (SDP <15) with FR = 0% was satisfying. Similarly, in the MRS of healthy other regions ($P_5$-$P_7$), Glx was also the bottom line (RE = 8.3 ± 0.9, FR = 5.7%) and the other metabolic quantifications achieved better RE ≤ 6.8 with FR ≤ 0.7%. SDP of all metabolic quantification was lower than 6.7 with FR = 0%. Finally, in the challenging lesion regions ($P_8$-$P_9$), tCho had the lowest RE=3.6±0.4 with FR=0% and tCr had RE=5.6±0.6 with FR=0.3%. Glx ( RE=9.8±1.1 with FR=10.0%, SDP=5.6±0.2 with FR=0%) was slightly worse and mI was the bottom line.

The denoised spectra by the ReLSTM also could provide comparable and credible relative concentration estimations for 5 major brain metabolic concentrations to 128 SA. Specifically, in relative concentration quantification, the RE did not exceed 11.0% (mI) and the FR (RE > 20) was no more than 11.1% of all metabolites. The minimum error could reach 3.6 ± 0.3 at the level of mmol/kg in the tCho. Specifically, in the MRS of the healthy frontal lobe ($P_1$-$P_4$), the most challenging case was Glx whose RE was 8.7 ± 0.8 with FR = 5.3%. The other metabolic quantifications achieved RE ≤ 5.5 with FR = 0%. In the MRS data of healthy other regions ($P_5$-$P_7$), mI was with the highest RE = 8.5 ± 1.0 and the corresponding FR = 7.7%, Glx had RE

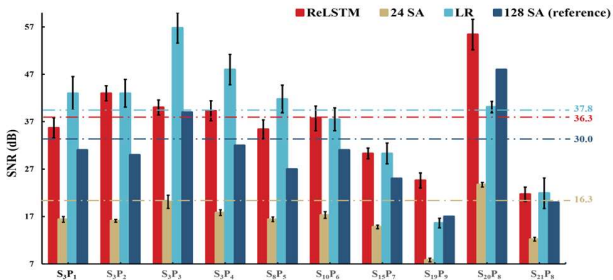

**Fig. 6.** SNR improvements of different denoised methods. Note: The dashed line indicated the mean value of SNR improvement.

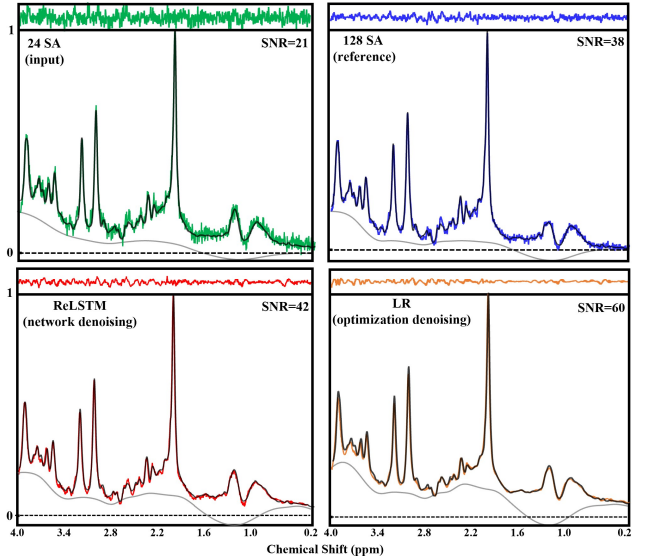

**Fig. 7.** The spectra of the healthy subject ($S_3P_3$) were denoised by four methods including the 24 SA (low-SNR input of the LR and ReLSTM), 128 SA, proposed ReLSTM and LR. Note: The LCModel fitting spectra and residuals (top) were presented with colour curves, and corresponding denoised spectra and fitted baseline were marked black and grey, respectively.

= 8.1 ± 0.9 with FR = 4.3%, and the other metabolic quantifications (tNAA and tCho) achieved RE ≤ 4.5 with FR = 0%. Finally, in the challenging lesion regions ($P_8$-$P_9$), tCr had the lowest RE = 4.8±0.5 with FR = 0% while Glx was slightly worse and mI was the bottom line.

In addition, The ReLSTM had a significant advantage over the LR in quantifying metabolites Glx and tCho no matter in healthy and lesion regions (significant lower quantification RE with lower FR in Glx, lower ER in tCho), which is shown in TABLE II. For example, in Glx relative quantification, the denoised spectra by the ReLSTM had significant lower RE and

**TABLE II**
Quantification relative error (RE) between 128 SA high SNR spectra and the 24 SA spectra denoised by the ReLSTM or LR.

| Position | Metabolite | Method | Absolute quantification | | | | Relative quantification | |
|---|---|---|---|---|---|---|---|---|
| | | | RE | FR (RE>20) | SDP | FR (SDP >15) | RE | FR (RE>20) |
| Health $P_1$-$P_4$ $N$=400 | tNAA | LR | **2.9±0.3** | 0% | 2.0±0.0 | 0% | **3.6±0.3** | 0% |
| | | ReLSTM | 3.4±0.3 | 0% | 2.0±0.0 | 0% | 4.0±0.4 | 0% |
| | tCr | LR | 2.6±0.2 | 0% | 1.6±0.0 | 0% | - | 0% |
| | | ReLSTM | 2.5±0.2 | 0% | 2.0±0.0 | 0% | - | 0% |
| | tCho | LR | 3.8±0.4 | 0% | 2.8±0.0 | 0% | 4.7±0.6 | 0% |
| | | ReLSTM | **3.0±0.3** | 0% | **2.2±0.1** | 0% | **3.2±0.3** | 0% |
| | Glx | LR | 26.1±1.6 | 70.0% | 7.1±0.1 | 0% | 25.1±1.3 | 69.0% |
| | | ReLSTM | **10.1±1.0** | **7.8%** | **6.4±0.1** | 0% | **8.7±0.8** | **5.3%** |
| | mI/Ins | LR | 5.6±0.5 | 0% | 4.0±0.0 | 0% | **4.9±0.5** | 0% |
| | | ReLSTM | **4.2±0.4** | 0% | 4.0±0.0 | 0% | 5.0±0.5 | 0% |
| Health $P_5$-$P_7$ $N$=300 | tNAA | LR | 3.2±0.4 | 0% | 1.9±0.0 | 0% | 3.7±0.4 | 0% |
| | | ReLSTM | 3.0±0.3 | 0% | 2.0±0.1 | 0% | 3.6±0.4 | 0% |
| | tCr | LR | 3.2±0.4 | 0% | 2.0±0.0 | 0% | - | 0% |
| | | ReLSTM | **2.8±0.3** | 0% | 2.0±0.0 | 0% | - | 0% |
| | tCho | LR | 3.8±0.5 | 0% | 2.8±0.1 | 0% | 4.7±0.6 | 0% |
| | | ReLSTM | 3.7±0.4 | 0% | 2.8±0.1 | 0% | **4.1±0.4** | 0% |
| | Glx | LR | 17.0±1.5 | 37.3% | 7.9±0.1 | 0% | 17.8±1.7 | 41.3% |
| | | ReLSTM | **8.3±0.9** | **5.7%** | **6.7±0.1** | 0% | **8.1±0.9** | **4.3%** |
| | mI/Ins | LR | 6.8±0.8 | 2.3% | 4.6±0.1 | 0% | 10.0±1.2 | 15.3% |
| | | ReLSTM | **6.1±0.7** | **0.7%** | 4.6±0.1 | 0% | **8.5±1.0** | 7.7% |
| Lesion $P_8$-$P_9$ $N$=300 | tCr | LR | 5.8±0.7 | 1.0% | 3.7±0.2 | 0% | 4.7±0.5 | 0% |
| | | ReLSTM | 5.6±0.6 | 0.3% | **3.4±0.2** | 0% | 4.8±0.5 | 0% |
| | tCho | LR | 3.9±0.4 | 0% | 2.2±0.1 | 0% | - | 0% |
| | | ReLSTM | **3.6±0.4** | 0% | **1.8±0.1** | 0% | - | 0% |
| | Glx | LR | 13.6±1.6 | 27.7% | 6.9±0.3 | 0% | 14.9±1.7 | 29.7% |
| | | ReLSTM | **9.8±1.1** | **10.0%** | **5.6±0.2** | 0% | **10.3±1.1** | **11.3%** |
| | mI/Ins | LR | **16.4±2.5** | 28.0% | 12.9±2.1 | 32.3% | **17.3±2.5** | 28.7% |
| | | ReLSTM | 17.4±2.8 | 28.3% | **10.5±1.8** | **22.3%** | 17.9±2.8 | 29.3% |
| Total $N$=1000 | tNAA** | LR | 3.0±0.2 | 0% | 1.9±0.0 | 0% | 3.6±0.3 | 0% |
| | | ReLSTM | 3.2±0.2 | 0% | 2.0±0.0 | 0% | 3.8±0.3 | 0% |
| | tCr | LR | 3.7±0.3 | 0.3% | 2.4±0.1 | 0% | 4.6±0.5 | 0%* |
| | | ReLSTM | 3.5±0.2 | 0.1% | 2.4±0.0 | 0% | 4.9±0.5 | 0%* |
| | tCho | LR | 4.2±0.2 | 0% | 2.3±0.0 | 0% | 4.2±0.3 | 0%** |
| | | ReLSTM | **3.4±0.2** | 0% | 2.3±0.0 | 0% | **3.6±0.3** | 0%** |
| | Glx | LR | 19.7±1.0 | 47.4% | 8.3±0.1 | 0% | 19.9±1.0 | 48.8% |
| | | ReLSTM | **9.5±0.5** | **7.8%** | **6.2±0.1** | 0% | **9.0±0.5** | **6.8%** |
| | mI/Ins | LR | 9.2±0.9 | 9.1% | 6.9±0.7 | 9.7% | 10.1±1.0 | 13.2% |
| | | ReLSTM | **8.7±1.0** | **8.7%** | **6.1±0.6** | **6.7%** | **10.0±1.0** | **11.1%** |

Note: The RE and SDP are applied the 99% confidence interval. we set RE≤20 and SDP≤15 as the criteria of acceptable reliability and counted the percentage of those which are exceeding criteria (failure rate (FR)). $N$=400, 300, and 300 are the numbers of the test sets from frontal lobe of healthy subject, the other three regions of healthy subjects, and patients' lesion regions, respectively. The "-" indicated the relative concentration was not evaluated because this metabolite was used as a reference, and the trace tNAA with the large SDP was also not evaluated in the lesion spectra. The * represented $N$=300 and ** represented $N$=700, respectively.

SDP for both healthy and lesion brain regions. The LR provided unreliable quantification (48.8% failed in total, 69.0% in the health $P_1$-$P_4$, 41.3% in the health $P_5$-$P_7$, 29.7% in the lesion $P_8$-$P_9$) since the failed proportion was very high. In the mI and tCho quantification, the denoised spectra by the ReLSTM also had lower RE and SDP for both healthy and lesion brain regions. Finally, for other metabolites, such as tNAA and tCr, the ReLSTM provided comparable quantification results to the LR in all testing regions. Thus, the overall quantification error (RE), SDP and FR of the ReLSTM is lower than that of the LR.

V. DISCUSSIONS

*A. Number of SA*

As seen in Fig.6 and TABLE II, selecting 24 SA ($m_1$=24) as input to train the ReLSTM, the SNR and quantification performances were comparable to 128 SA. The ReLSTM also outperformed LR in the quantification of Glx, tCho and mI for all data, except for one patient case $S_{20}P_8$ that had a failure percentage of more than 60% in mI quantification (Supplement TABLE S1). The reason is that, even for the maximal average ($m_1$=128), the uncertainty of mI quantification is very high (SDP>15%). Then, both LR and ReLSTM methods had a large failure percentage of over 60% (Supplement TABLE. S1-2).

Fewer repeated samples ($m_1$<24) for SA shows significantly worse performance in our testing set (Supplement TABLE. S3). A lower SNR spectrum (<24 SA) is far from the corresponding high SNR 128 SA spectrum in terms of peak shapes, noise intensity and background signals (Fig. 8(a)). Such big differences would pose great challenges to denoising. For example, with 8 SA, unwanted signals and poor recovery of small peaks remained after the ReLSTM or LR denoising (Fig. 8(d) and (g)). These problems can be improved with the SNR improvement of the input (Fig. 8(c), (f) and (i)) but these are not efficient for time-saving. Therefore, we suggested choosing $m_1 \geq 24$ to have a reasonable SNR in inputs (Fig.6) and set $m_1$=24 to save the scanning time as much as possible (80%).

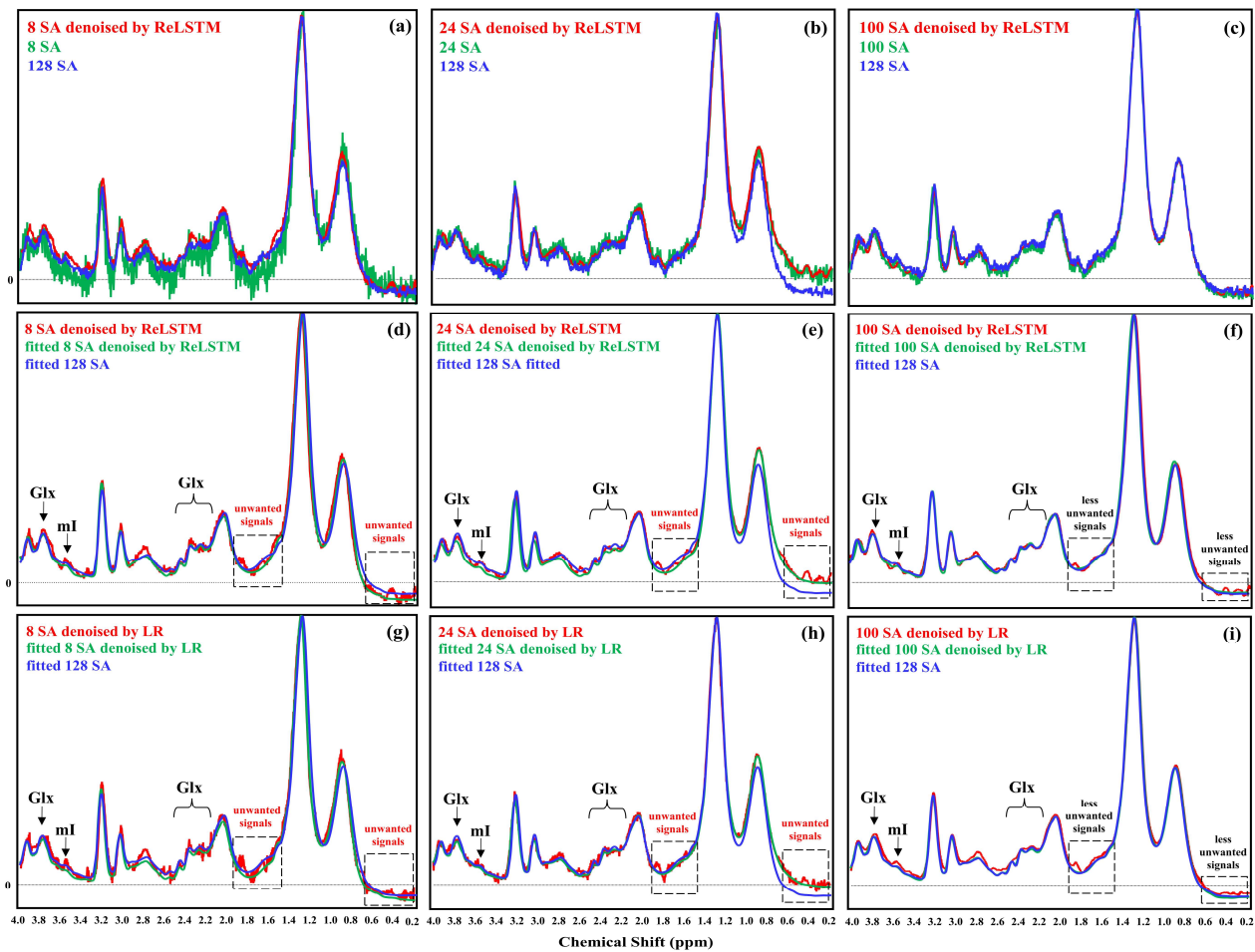

**Fig. 8.** Different SA inputs on the ReLSTM or LR denoising, and the corresponding LCModel spectrum fittings (taking a patient $S_{20}P_8$ for example). (a-c) are the comparison of different SA spectra and the ReLSTM denoised spectra; (d-f) are the comparison of the ReLSTM denoised spectra and corresponding fitted spectra; (g-i) are the comparison of the LR denoised spectra and corresponding fitted spectra; The 128 SA and corresponding fitted spectra was set as the reference. As can be seen, the LR and ReLSTM denoising results got better (especially on the small peaks Glx), as the SA of the input increased.

In addition, to enhance the diversity of labels, we chose $m_2$=124, and the theoretical SNR enhancement of 124 is very close to that of 128 (SNR improvement factor $\sqrt{124} \approx \sqrt{128}$ ).

*B. Unwanted Signals*

Any poor-quality realistic input may carry some unwanted signals. Neither LR nor ReLSTM denoising can remove these unwanted signals introduced by poor input without any prior knowledge. Fig.8 (e) and (h) show an example that spectra denoised by the ReLSTM or LR remain the unwanted signals introduced by poor input (green in Fig.8 (b)) at about 0.2-0.7 ppm and 1.4-1.9 ppm. Unwanted signals could be reduced if the SA increases (Fig.8 (f) and (i)).

Even though, only a mild impact was made by unwanted signals on quantification. The spectrum fitting, such as with LCModel, has the tolerance to these unwanted signals because these unwanted peaks are not in the predefined basis-set, which are shown as green lines in Fig.8 (d)-(i).

*C. Feasibility of the ReLSTM*

This deep learning denoising method showed potential feasibility in a wide range of clinical significance voxel positions. For example, the ReLSTM denoising model was trained using only the measured spectra from 25 healthy frontal regions and it could apply to other subjects' frontal regions or other brain regions including the occipital lobe, parietal lobe and posterior cingulate cortex, even in the lesion region spectra in our test. This potential feasibility needs further verification on more MRS data, which would be interesting for future work.

## VI. CONCLUSIONS

In this work, we proposed a deep learning denoising scheme for *in vivo* magnetic resonance spectroscopy. The network was solely trained on the *in vivo* MRS data with augmentation by recombining repeated samples. The denoising performance on data with reduced sampling (24 SA), including relative error and standard deviations percentage in quantifying some important biomarkers (tNAA, tCho, tCr, Glx and mI), were comparable with a high number of averages (128 SA). This means that the proposed method may save about 80% of data acquisition time and enable fast single-voxel MRS. Although the network training was performed on healthy volunteer data, the trained network also showed feasibility on patient data that had cerebral infarction and tumor regions in brains, indicating promising applications in clinics.

## ACKNOWLEDGEMENTS

The authors thank all of the volunteers for assisting in data acquisition and also thank the staff from the Zhongshan Hospital affiliated to Xiamen University, the Second Affiliated Hospital of Xiamen Medical College, scientists from Philips Healthcare and the Shanghai Jiao Tong University for technical support. The authors also sincerely appreciate editors and reviewers for their constructive suggestions that improve this work.